\documentclass[preprint,number]{elsarticle}
\usepackage{dcolumn}
\usepackage{graphicx,color}
\usepackage{textcomp}
\begin {document}
  \newcommand {\nc} {\newcommand}
  \nc {\beq} {\begin{eqnarray}}
  \nc {\eeq} {\nonumber \end{eqnarray}}
  \nc {\eeqn}[1] {\label {#1} \end{eqnarray}}
  \nc {\eol} {\nonumber \\}
  \nc {\eoln}[1] {\label {#1} \\}
  \nc {\ve} [1] {\mbox{\boldmath $#1$}}
  \nc {\ves} [1] {\mbox{\boldmath ${\scriptstyle #1}$}}
  \nc {\mrm} [1] {\mathrm{#1}}
  \nc {\half} {\mbox{$\frac{1}{2}$}}
  \nc {\thal} {\mbox{$\frac{3}{2}$}}
  \nc {\fial} {\mbox{$\frac{5}{2}$}}
  \nc {\la} {\mbox{$\langle$}}
  \nc {\ra} {\mbox{$\rangle$}}
  \nc {\etal} {\emph{et al.}}
  \nc {\eq} [1] {(\ref{#1})}
  \nc {\Eq} [1] {Eq.~(\ref{#1})}
  \nc {\Ref} [1] {Ref.~\cite{#1}}
  \nc {\Refc} [2] {Refs.~\cite[#1]{#2}}
  \nc {\Sec} [1] {Sec.~\ref{#1}}
  \nc {\chap} [1] {Chapter~\ref{#1}}
  \nc {\anx} [1] {Appendix~\ref{#1}}
  \nc {\tbl} [1] {Table~\ref{#1}}
  \nc {\fig} [1] {Fig.~\ref{#1}}
  \nc {\ex} [1] {$^{#1}$}
  \nc {\Sch} {Schr\"odinger }
  \nc {\flim} [2] {\mathop{\longrightarrow}\limits_{{#1}\rightarrow{#2}}}
  \nc {\textdegr}{$^{\circ}$}
  \nc {\IR} [1]{\textcolor{red}{#1}}
  \nc {\IG} [1]{\textcolor{green}{#1}}
\title{One-neutron halo structure by the ratio method}
\author[msu,him]{P.~Capel\corref{cor}}
\ead{pierre.capel@centraliens.net}
\author[surrey,msu]{R.~C.~Johnson}
\ead{r.johnson@surrey.ac.uk}
\author[msu]{F.~M.~Nunes}
\ead{nunes@nscl.msu.edu}

\address[msu]{National Superconducting Cyclotron Laboratory and Department of Physics and Astronomy, Michigan State University, East Lansing, Michigan 48824, USA}
\address[him]{Helmholtz-Insitut Mainz, Johannes Gutenberg-Universit\"at, D-55128 Mainz, Germany}
\address[surrey]{Department of Physics, University of Surrey, Guildford GU2 7XH, United Kingdom}

\cortext[cor]{Corresponding author}

\begin{abstract}
We present a new observable to study halo nuclei.
This new observable is a particular \emph{ratio} of angular
distributions for elastic breakup and scattering.
For one-neutron halo nuclei, it is shown to be independent of the reaction mechanism
and to provide significant information about the structure
of the projectile, including
binding energy, partial-wave configuration, and radial wave function
of the halo.
This observable offers new capabilities for the study of
nuclear structure far from stability.
\end{abstract}
\begin{keyword}
Halo nuclei \sep angular distribution \sep elastic scattering \sep breakup
\end{keyword}
\maketitle
%

%%%%%%%%%%%%%%%%%%%%%%%%%%%%%%%%%%%%%%%%%%%%%%%%%%%%%%%%%%%%%%%%%%%%%%%%%%%%%

Nuclear halos are one of the most striking phenomena revealed 
through the study of extreme states of matter. 
Their discovery became possible through the development of 
radioactive beams in the mid 80s \cite{Tan85r}. 
Measured reaction cross sections along an isotopic chain are 
seen to increase dramatically as the limits of stability are 
approached as compared with more bound neighboring isotopes. 
This observation implies that near the end of an isotopic chain 
(the drip-line), where the neutron number is much larger 
than the proton number, adding one or two neutrons to a 
well-bound core may produce a nucleus with a radius much larger 
than that of its core, suggesting a  halo picture for 
these valence neutrons \cite{HJ87,Tan96}.

The halo phenomenon is believed to arise from the combination of 
a very low separation energy of the valence particles and 
the absence of a strong repulsive barrier.
For example, this may appear for neutrons loosely bound
to a core in an $s$ orbital.
The result is a highly delocalized wave function and a considerable 
probability of finding the valence particles
outside the range of their binding
interaction to the core,
well into the classically forbidden region. 
It is less likely to find nuclear halos on the 
proton drip-line or in orbitals involving large angular momentum, 
as the Coulomb or centrifugal barriers hinder the development
of the extended wave function. 
Halo structures have
also been observed experimentally in other fields such as atomic 
and molecular  physics \cite{JRF04}.
Extreme examples of halo states are Efimov states, an area generating 
intense activity in molecular physics \cite{Zac09}. 
 
These qualitative features of halo states are fairly well established, 
but ever since the discovery of halo nuclei \cite{Tan85r},
it has been a challenge to reconcile
this picture with the strongly interacting many-fermion 
structure of real nuclei.
On the experimental side much of 
the difficulty arises because halo nuclei tend to be unstable 
against the weak interaction and therefore cannot be prepared as a target. 
Plans are being made to use electron scattering from 
trapped radioactive ions \cite{Sud09} or from
a beam \cite{Ant11} of these
exotic species to measure their charge distribution.
So far all the available evidence for their structure is 
obtained indirectly mostly through the measurements of nuclear
reactions, such as breakup \cite{Nak06}, elastic scattering \cite{Dip09l},
or knockout \cite{Nak09l}.

Collisions between a one-neutron halo nucleus and a target
can be described as three-body processes in which a projectile $P$,
seen as a valence neutron n loosely bound to a core $c$,
impinges on a target $T$.
Various theoretical models have been developed to solve the corresponding
\Sch equation (see \Ref{Bay08} for a review).
These models have improved our understanding of the reaction process.
They have shown that the mechanisms involved in
collisions of loosely-bound nuclei are more complex than initially thought
and that extracting information about nuclear structure
from reaction measurements is not as straightforward as hoped \cite{CB05}.
Moreover, reaction calculations depend on
optical potentials, which describe the interaction between the projectile
constituents and the target. These potentials are often unknown.
This is especially true
for the interaction of the core and the target as there exist little---if any---data to constrain this potential,
the core being usually itself radioactive.
The uncertainties related to the choice of these potentials
hinder the quantitative analysis of experimental data \cite{CGB04}.
An observable that is less dependent on the reaction process and
that reveals more information about the structure of the projectile is
clearly needed. In this Letter, we present such an observable.

In this framework, the structure of the projectile is described
by the internal Hamiltonian
\beq
H_0=\frac{p^2}{2\mu_{c\rm n}}+V_{c\rm n}(\ve{r}),
\eeqn{ea}
where $\mu_{c\rm n}$ is the $c$-n reduced mass,
$\ve{r}$ is the relative $c$-n coordinate,
and $\ve{p}$ the corresponding momentum.
The $c$-n potential $V_{c\rm n}$ is adjusted to reproduce
properties of the projectile, such as its binding energy and
some of its excited levels.
In partial wave $ljm$, the eigenstates of $H_0$ of energy $E$
are denoted by $\phi_{ljm}(E)$
($l$ is the $c$-n orbital angular momentum,
$j$ is the total angular momentum resulting from the coupling of $l$ with the
neutron spin, and $m$ is the projection of $j$).
For $E<0$ they are normed to unity and correspond to bound states of the projectile.
For $E>0$, they describe the continuum spectrum,
i.e. the projectile broken up into $c$ and n.
They are normalized as
$\langle\phi_{ljm}(E)\mid \phi_{l'j'm'}(E^{\prime})\rangle
=\delta_{ll'}\delta_{jj'}\delta_{mm'}\delta(E-E^{\prime})$.
The interactions of the projectile fragments $c$ and n with the target
are simulated by the optical potentials $V_{cT}$ and $V_{{\rm n}T}$, respectively.
Within this framework the study of reactions involving one-neutron halo nuclei
reduces to solving the three-body \Sch equation with Hamiltonian
\beq
H=\frac{P^2}{2\mu}+H_0
+V_{cT}\left(\ve{R}-\frac{m_{\rm n}}{m_P}\ve{r}\right)
+V_{{\rm n}T}\left(\ve{R}+\frac{m_c}{m_P}\ve{r}\right),
\eeqn{eb}
where $\mu$ is the $P$-$T$ reduced mass,
$m_{\rm n}$ is the mass of the valence neutron, and $m_c$ that of the core
($m_P=m_c+m_{\rm n}$).
Variable $\ve{R}$ is the $P$-$T$ relative coordinate and $\ve{P}$
the corresponding momentum.
The \Sch equation corresponding to Hamiltonian \eq{eb} must
be solved with the condition that the impinging projectile
is initially in its ground state $\phi_0$.

Recently, within the dynamical eikonal approximation (DEA) \cite{BCG05,GBC06,Bay08},
angular distributions for the elastic scattering
and elastic breakup of one-neutron halo nuclei have been studied \cite{CHB10}.
This analysis shows that both processes exhibit very similar features,
suggesting that the loosely-bound projectile is scattered similarly
whether it remains in its ground state or is broken up.
This result can be explained within the
Recoil Excitation and Breakup (REB) model \cite{JAT97,Joh99}.
In this model, a simple solution of the three-body \Sch equation
is obtained by neglecting $V_{{\rm n}T}$ and
the excitation energy of the projectile
(i.e. using the adiabatic---or sudden---approximation).
In the REB limit the elastic-scattering cross section in direction
$\Omega=(\theta,\phi)$ in the $P$-$T$ center-of-mass rest frame
is exactly factorized into the product of an elastic-scattering cross section
for a pointlike projectile $(d\sigma/d\Omega)_{\rm pt}$
and a form factor describing the extension
of the halo \cite{JAT97,Joh99}.

The REB model can also describe the angular distributions
for excitation of the projectile
to any of its states \cite{Joh99,Joh98}.
The corresponding cross sections also factorize
into a reaction-dynamics part and a projectile-structure part.
In particular, this can be performed for the breakup of the projectile,
i.e. its excitation at an energy $E$ in the $c$-n continuum with its center
of mass scattered in direction $\Omega$.
To the extent that the small difference in magnitude between
the outgoing momenta for elastic scattering and
breakup can be neglected, the point-like cross section $(d\sigma/d\Omega)_{\rm pt}$
is identical in the expression of both processes.
This particular feature explains why the angular distributions for
elastic scattering and breakup are so similar \cite{CHB10}.
It also leads to the main new idea introduced here.
It is exactly $(d\sigma/d\Omega)_{\rm pt}$ that contains
the undesired dependence on the $P$-$T$ relative motion
and its sensitivity to $V_{cT}$.
Therefore, a ratio of breakup to elastic-scattering angular distributions
would naturally remove this dependence and
provide information pertaining only to the halo structure.
In addition, this observable, being the ratio of two cross sections,
would not depend on their absolute normalizations,
which is particularly attractive from an experimental point of view.

The same factorization is obtained for the ratio of
any linear combination of angular distributions.
We have found it optimal to consider the ratio of
the angular distribution for elastic breakup at one definite $c$-n
relative energy $E$, to the sum of the angular distributions
for elastic and inelastic scattering and for
elastic breakup at all $c$-n energies
\beq
\frac{d\sigma_{\rm sum}}{d\Omega}=\frac{d\sigma_{\rm el}}{d\Omega}
+\frac{d\sigma_{\rm inel}}{d\Omega}
+\int \frac{d^2\sigma_{\rm bu}}{dEd\Omega} dE.
\eeqn{e6}
Using the closure relation for the states of the projectile,
this ratio is approximated at the REB limit by
\beq
\left(\frac{d^2\sigma_{\rm bu}/dEd\Omega}{d\sigma_{\rm sum}/d\Omega}\right)_{\rm REB}=
|F_{E,0}(\ve{Q})|^2,
\eeqn{e5}
where the form factor reads
\beq
|F_{E,0}|^2=\sum_{ljm}\left|\int\phi_{ljm}(E,\ve{r}) \phi_0(\ve{r})
e^{i\ve{Q\cdot r}}d\ve{r}\right|^2,
\eeqn{e4}
with $\ve{Q}=(m_{\rm n}/m_P)(\ve{K}-\ve{K'})$
corresponding to the fraction $m_{\rm n}/m_P$ of the momentum
transferred from the incoming $\hbar\ve{K}$ to the
outgoing $\hbar\ve{K'}$ momenta between the center of mass of the $c$-n pair
and the target.
We show below by comparison with more exact calculations that
the REB approximations can be justified in realistic cases.
Assuming this temporarily we discuss the structure
information that can be extracted from $|F_{E,0}|^2$. 

\begin{figure}
\center
\includegraphics[width=8cm]{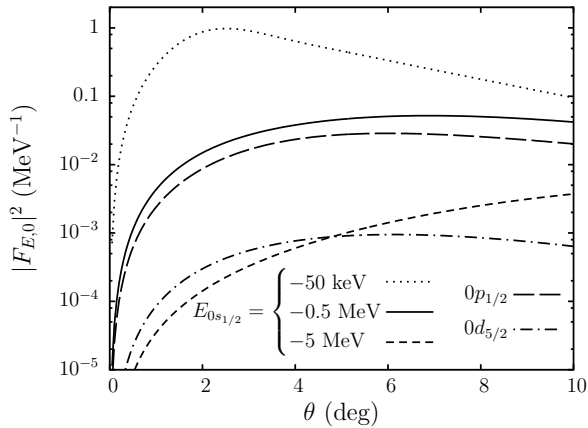}
\caption{Form factor $|F_{E,0}|^2$ for \ex{11}Be impinging on Pb
at 69~MeV/nucleon.
Its sensitivity to the projectile binding energy and
partial-wave configuration is illustrated.
}\label{f2a}
\end{figure}

\begin{figure}
\center
\includegraphics[width=8cm]{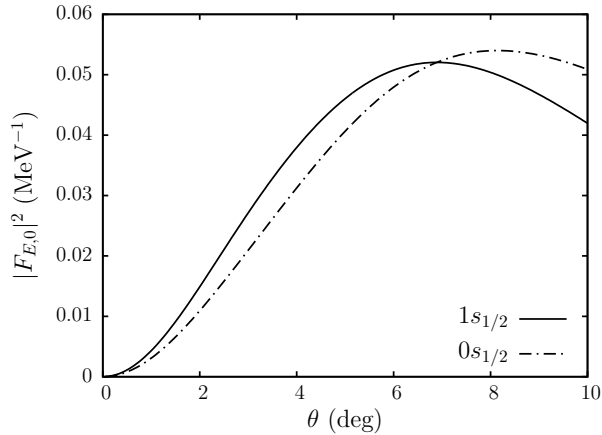}
\caption{Sensitivity of the form factor $|F_{E,0}|^2$ to changes in
the radial $c$-n wave function.
Differences are observed between (realistic) $1s_{1/2}$
and $0s_{1/2}$ ground states.
%for \ex{11}Be impinging on Pb at 69~MeV/nucleon.
}
\label{f1}
\end{figure}

To illustrate the general properties of $|F_{E,0}|^2$, we consider
\ex{11}Be as a projectile.
Reactions involving this archetypal one-neutron halo nucleus
have been extensively studied both
theoretically and experimentally.
In particular its angular distributions for breakup
on Pb and C targets have been precisely measured at about
70~MeV/nucleon \cite{Fuk04}.
Figs.~\ref{f2a} and \ref{f1} depict $|F_{E,0}|^2$ at $E=0.1$~MeV
for \ex{11}Be impinging on lead at 69~MeV/nucleon as a function
of the azimuthal angle $\theta$ up to $10^\circ$.
The form factor \eq{e4} is initially computed for the $c$-n potential developed
in \Ref{CGB04}, in which the $1/2^+$ ground state of \ex{11}Be is produced
in the $1s_{1/2}$ orbital (solid line in both Figs.~\ref{f2a} and \ref{f1}).
%, i.e. with one node in the interior (solid line).
%Performing a series of tests, we find that the shape and magnitude
%of $|F_{E,0}|^2$ depend strongly on
%the angular momentum of the valence neutron
%and its binding energy to the core.
%The major interest of this ratio method is its strong dependence on
%the projectile internal structure.h
The dependence of this form factor on the projectile internal structure
is then analyzed.
First, in \fig{f2a}, the sensitivity of $|F_{E,0}|^2$ to the
$c$-n binding energy is
investigated by varying the $s$-wave depth of $V_{c\rm n}$ to reproduce
the $1s_{1/2}$ ground state at 50~keV (dotted line),
the realistic 0.5~MeV (solid line), and 5~MeV (short-dashed line).
The dependence on the partial-wave configuration of
the ground state is also analyzed in \fig{f2a}
by fitting the 0.5-MeV binding energy
in the $0p_{1/2}$ (long-dashed line) and $0d_{5/2}$ (dash-dotted line)
orbitals.
We observe that both shape and magnitude of the form factor depend strongly
on the angular momentum of the valence neutron
and its binding energy to the core.
This shows that the ratio method could be a very useful tool
to measure precisely and simultaneously these two quantities for
loosely-bound nuclei, which,
such as \ex{31}Ne \cite{Nak09l},
are too short-lived for usual spectroscopic techniques.

Interestingly, the form factor varies also with the shape of
the radial part of the ground-state wave function $\phi_0$.
This point is illustrated in \fig{f1},
where the dash-dotted line corresponds to $|F_{E,0}|^2$
obtained from the $V_{c\rm n}$ of \Ref{CGB04} but with
a depth in the $s_{1/2}$ partial wave reduced to produce
the ground state of \ex{11}Be in the $0s_{1/2}$ orbital, i.e. without node.
We observe interesting changes in the form factor.
At forward angles, $|F_{E,0}|^2$ for the $1s_{1/2}$ bound state
is larger than for the $0s_{1/2}$ one.
On the contrary, at larger angles,
the $0s_{1/2}$ form factor exceeds the $1s_{1/2}$ one.
This suggests that depending on the angle considered, $|F_{E,0}|^2$
probes different parts of the radial wave function.
Up to $2^\circ$ the form factor scales with the square of the 
asymptotic normalization coefficient (ANC)
that determines the asymptotic exponential form of  $\phi_0$. %,
The crossing of the form factors shows that larger
angles probe the interior of $\phi_0$, as in that region the $0s_{1/2}$
wave function, exhibiting no node, is larger than the $1s_{1/2}$ one.
Albeit small, the variations observed between both form factors
show that the ratio method probes a wider range of the halo wave function
than usual breakup observables, which are purely peripheral \cite{CN07}.

The richness of halo-structure information residing in 
$|F_{E,0}|^2$ makes the ratio idea particularly attractive.
We now address the key question of the accuracy
in realistic cases of the REB approximation to the three-body
model on which \Eq{e5} is based. 
If the adiabatic treatment of the internal motion of the
projectile is lifted and the n-$T$ interaction
is included, does \Eq{e5} hold?
To answer this question, we make use of the DEA \cite{BCG05,GBC06}
which reproduces closely the data of \Ref{Fuk04} and does not
make the approximations introduced in the REB model.
In this analysis,
we use the inputs and numerical conditions detailed in \Ref{GBC06}.
%%%%%%%%%%%%%%%%%%%%%%%%%%%%%%%%%%%%%%%%%%%%%%%%%%%%%%%%%%%%%%%%%%%%%%%%%

\begin{figure}
\center
\includegraphics[width=8cm]{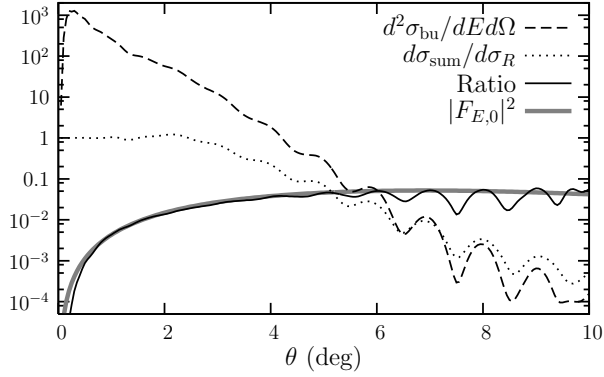}
\caption{Ratio of the breakup and
summed angular distributions for \ex{11}Be on Pb at 69~MeV/nucleon.
DEA calculation is compared to its REB estimation $|F_{E,0}|^2$.}
\label{f2}
\end{figure}

In \fig{f2} we present the DEA angular distributions for \ex{11}Be
impinging on lead at 69~MeV/nucleon.
The breakup angular distribution is displayed for a
relative $c$-n energy $E=0.1$ MeV in b/MeV/sr (dashed line).
The summed angular distribution \eq{e6} is shown as
a ratio to Rutherford (dotted line).
As observed in \Ref{CHB10}, both exhibit very similar features
(Coulomb rainbow, oscillatory patterns,\ldots).
In the ratio of these distributions (thin solid line),
these features are smoothed out, confirming that most
of the dependence on the reaction dynamics is removed
by this technique.
Moreover, since the DEA ratio compares very well with its REB prediction
$|F_{E,0}|^2$ (gray thick line), we conclude that the ratio method
gives direct access to halo-structure information.

Even though the agreement is generally good, we observe
discrepancies between the DEA ratio and $|F_{E,0}|^2$. In particular,
at large angles
the former exhibits small oscillations whereas the latter is smooth.
To investigate the source of these differences,
we repeat the DEA calculation setting $V_{{\rm n}T}=0$.
We find that the oscillations in the ratio
disappear and the agreement with  $|F_{E,0}|^2$ improves.
As shown by Johnson \etal, $V_{{\rm n}T}$ shifts the oscillatory
pattern of angular distributions \cite{JAT97}.
This shift, depending on the excitation energy of the projectile,
brings the breakup and scattering distributions
slightly out of phase, causing the oscillations
of the ratio in the realistic case.
This result indicates that $V_{{\rm n}T}$ has non-negligible
effects on reaction calculations.
However, these effects remain small compared to the
dynamical effects removed by the ratio \eq{e5}.

\begin{figure}
\center
\includegraphics[width=8cm]{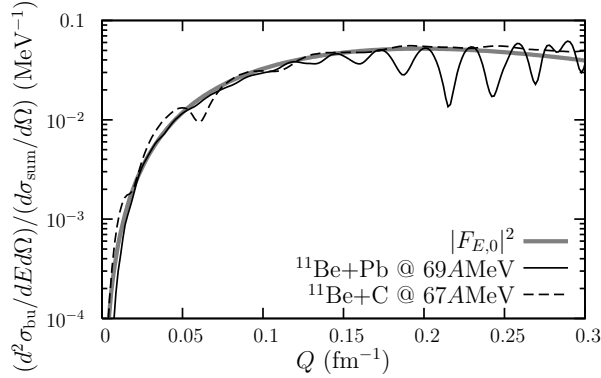}
\caption{Sensitivity of ratio \eq{e5} to the reaction mechanism.
DEA calculations on different targets are compared to $|F_{E,0}|^2$.}
\label{f3}
\end{figure}

To evaluate the independence to the $c$-$T$ interaction gained by the ratio
method, we repeat the DEA calculation with different choices
of $V_{cT}$. Switching off the nuclear part of that optical
potential does not lead to significant changes
in the ratio \eq{e5}, although the characteristics of the angular
distributions
are significantly altered \cite{CHB10}.
The ratio removes the dependence on this interaction,
which is a large source of uncertainty in the analysis
of experimental data.
This result suggests the ratio to be independent of the target,
and hence of the reaction mechanism.
To confirm this, we compare in \fig{f3} the results on Pb
at 69~MeV/nucleon (thin solid line) with those on C
at 67~MeV/nucleon (dashed line).
Now the ratio is plotted as a function of $Q$
to allow the comparison between different targets.
Although the reaction mechanism on C is very different
from the Coulomb-dominated collision on Pb, the ratio for \ex{11}Be+C
is very close to the ratio for \ex{11}Be+Pb.
Both are in excellent agreement with $|F_{E,0}|^2$ (thick gray line).
This result shows that besides removing the uncertainty caused by
the projectile-target potentials, observable \eq{e5} is,
for all practical purposes, independent of the reaction mechanism.

%%%%%%%%%%%%%%%%%%%%%%%%%%%%%%%%%%%%%%%%%%%%%%%%%%%%%%%%%%%%%%%%%%%

In this Letter, we propose the study of the structure of halo nuclei
through a new observable consisting in the ratio of two angular distributions:
one for the elastic breakup of the projectile at a definite $c$-n energy
and the other corresponding to the elastic scattering and elastic breakup
of the projectile into any of its states.

We show that this ratio removes the dependence on the reaction mechanism and
leads to an observable that can provide detailed information on the halo
structure.
Besides the binding energy and the angular momentum of the valence neutron,
the radial wave function of the halo can be accessed by this ratio.
Depending on the angle considered, both asymptotic
and internal parts of the wave function can be probed.
To assess the influence of a multiconfiguration structure of the
projectile on this observable, a generalization of the REB model
is needed. This will be required to evaluate how
spectroscopic factors different from 1 affect the ratio.
A discussion on this matter and a detailed study of the
sensitivity of the ratio method to the projectile
description will be published elsewhere.
Derived here for one-neutron halo nuclei, this technique is most likely
extendible to multibody systems, such as two-neutron halo nuclei or
Efimov states in atomic and molecular physics \cite{JRF04,Zac09}.
An extension to charged valence particles may also be possible \cite{Hor10},
which would enable the study of proton halos and halo-like negative ions.
This observable looks therefore very promising for the study of
loosely-bound quantal structures.
At least in the realm of nuclear physics, we believe the ratio method will open
a new era in the study of exotic systems.

\section*{Acknowledgments}
We thank I.~J.~Thompson, B.~Tsang, N.~Timofeyuk, and the MoNA collaboration for interesting discussions on the subject.
This work was supported by the National Science Foundation grant PHY-0800026 and the Department of Energy under contract DE-FG52-08NA28552
and DE-SC0004087. RCJ is supported by the United Kingdom Science and Technology Facilities Council under 
Grant No. ST/F012012.

\end{document}